\documentclass[aps,prb,preprint,groupedaddress]{revtex4}

\usepackage{amsmath}
\usepackage{bm}
\usepackage{graphicx,subfigure}

\newcommand{\ket}[1]{\mid \! #1 \rangle}

\begin{document}

\title{Herzberg Circuit and Berry's Phase in Chirality-based Coded
Qubit in a Triangular Triple Quantum Dot}

\author{Chang-Yu Hsieh} 
\affiliation{Quantum Theory Group, 
Institute for Microstructural Sciences, 
National Research Council, Ottawa, Canada K1A 0R6} 
\affiliation{Department of Physics, 
University of Ottawa, Ottawa, ON, Canada, K1N 6N5} 

\author{Alexandre Rene} 
\affiliation{Quantum Theory Group, 
Institute for Microstructural Sciences, 
National Research Council, Ottawa, Canada K1A 0R6} 
\affiliation{Department of Physics, 
University of Ottawa, Ottawa, ON, Canada, K1N 6N5}

\author{Pawel Hawrylak} 
\affiliation{Quantum Theory Group, 
Institute for Microstructural Sciences, 
National Research Council, Ottawa, Canada K1A 0R6} 
\affiliation{Department of Physics, 
University of Ottawa, Ottawa, ON, Canada, K1N 6N5} 

\begin{abstract}
We present a theoretical proposal for the  Herzberg circuit and controlled accumulation of Berry's phase in a chirality-based coded qubit in a triangular triple quantum dot molecule with one electron spin each. The qubit is encoded in the two degenerate states of a three spin complex with total spin $S=1/2$. Using a Hubbard and Heisenberg model the Herzberg circuit encircling the degeneracy point is realized by adiabatically tuning the successive on-site energies of quantum dots and tunnel couplings across a pair of neighbouring dots.  It is explicitly shown that encircling the degeneracy point leads to the accumulation of the geometrical Berrys phase. We show that only triangular but not linear quantum dot molecule allows for the generation of Berry's phase and we discuss a protocol to detect this geometrical phase. 
\end{abstract}


\maketitle

\section{Introduction}
As discussed by Herzberg \cite{herzberg_longuethiggins_df1963} and Berry  \cite{berry_frs_rslA1984}, the wavefunction acquires geometric phase\cite{wilczek_shapere_book}, Berry's phase, when it is adiabatically moved along a circuit in parameter space of the Hamiltonian, the Herzberg circuit, enclosing a degeneracy point.  Since only the topology of the circuit determines whether the geometrical phase is accumulated, the Berry's phase is insensitive to the effects of interactions between the system and its environment. For this reason there is interest in attempting to  encode and manipulate quantum information in geometric phases, for example holonomic quantum computing \cite{carollo_vlatko_book2008} with generalized non-Abelian geometric phase \cite{pachos_zanardi_pra2000}. 

Experimentally, Berry's phase in two level systems has already been demonstrated, including  neutron \cite{bitter_dubbers_prl1987} and nuclear spins \cite{jones_vedral_nat2000}, superconducting qubits \cite{leek_flink_sci2007} and a superconducting charge pump \cite{mottonen_vartiainen_prl2008}.  Preceding the successful experiments with superconducting circuits,  the relation between Berry's phase of a superconducting circuit and other measurable quantities was investigated theoretically  \cite{aunola_toppari_prb2003,pekola_toppari_prb1999}, including a theoretical proposal  for realizing geometric quantum computation with superconducting qubits\cite{falci_fazio_nat2000}.  

In this work, we demonstrate theoretically the generation of Herzberg circuit and Berry's phase in quantum states of a three electron complex  in a triangular triple quantum dot molecule with one electron spin each within the framework of Hubbard and Heisenberg models. The two level system, a qubit, is encoded in the two degenerate states of a three spin complex with total spin $S=1/2$\cite{hawrylak_korkusinski_ssc2005, hsieh_hawrylak_prb2010,divincenzo_bacon_nat2000}.  An early proposal to generate geometrical phase in degenerate one electron quantum levels of a three-atom system was discussed by Herzberg and Longuet-Higgins in Ref.~\onlinecite{herzberg_longuethiggins_df1963} in 1963.   A  triple quantum dot (TQD) molecule studied here is related to the three-atom system.  We define a two-level system, coded qubit, \cite{hawrylak_korkusinski_ssc2005, hsieh_hawrylak_prb2010} by the two lowest degenerate  levels of a half-filled three electron TQD under an in-plane magnetic field.  It was shown that the quantum states of coded qubit in a TQD  can be  manipulated by tuning the gate voltages\cite{hawrylak_korkusinski_ssc2005, hsieh_hawrylak_prb2010}.   This opens the possibility described in this work   to engineer the Hamiltonian to undergo adiabatic and cyclic evolution along the Herzberg circuit resulting in accumulation of Berry's phase.   Recent experiments \cite{gaudreau_kam_apl2009,granger_gaudreau_prb2010,laird_taylor_prb2010} on the linear TQD have already demonstrated the high tunability and coherent manipulation of the many-body quantum states in a TQD.
However, we show that it is not possible to generate a Herzberg circuit for a linear triple quantum dot, only triangular triple quantum dot molecule with control over quantum dot energies and at least one tunneling amplitude is needed. 

The plan of the paper is as follows.  In Sec.\ref{sec:model}, we describe the system and the Hamiltonians for a triangular and linear TQD.  In Sec.\ref{sec:generation},  we construct Herzberg circuit generating Berry's phase in a triangular TQD.  We show that it is not possible to construct Herzberg circuit for a linear TQD.  In Sec.\ref{sec:conclusion}, a brief conclusion is given.

\section{The model} \label{sec:model}

\begin{figure}
  \subfigure[]{
     \includegraphics[width=60mm]{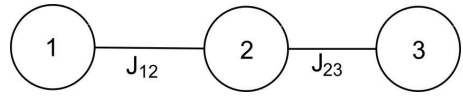}
	 \label{fig:triangleTQD}
  }

  \subfigure[]{
    \includegraphics[width=60mm]{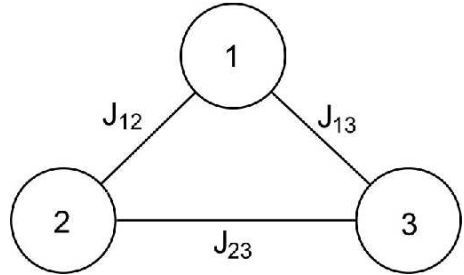}
	 \label{fig:linearTQD}
  }
  \caption{\label{fig:TQD} \subref{fig:triangleTQD} Schematic representation of a triangular triple quantum dot.\subref{fig:linearTQD} Schematic representation of a linear triple quantum dot.}
\end{figure}

A lateral TQD is defined by metallic gates on top of a two-dimensional electron gas in the $(x,y)$ plane at GaAs/AlGaAs heterojunction with three local minima, capable of confining a controlled number of electrons.  With one electron in each dot, the extended Hubbard Hamiltonian reads,
\begin{align}
\label{eq1_Hubbard}
\hat{H}_{hubb}= \sum_{i=1}^{3} E_{i} \hat{n}_{i\sigma} + \sum_{\substack{i, j = 1\\ i \neq j }}^{3} \sum_ \sigma t_{ij}(\mathbf{B}) \hat{c}^{\dag}_{i\sigma}\hat{c}_{j\sigma}
 +  \frac{1}{2} \sum_{\substack{i, j =1\\ i \neq j }}^{3}V_{ij} \hat{\rho}_{i} \hat{\rho}_{j} + \sum_{i=1}^{3}U_i \hat{n}_{i\uparrow}\hat{n}_{j\downarrow}+\sum_\alpha g\mu_B\mathbf{S}_\alpha \cdot \mathbf{B},
\end{align} 
where $\hat{\rho}_i = \hat{n}_{i\uparrow} + \hat{n}_{i\downarrow}$,  $E_i$ is the on-site energy, $t_{ij}$ is the tunnel coupling between dot $i$ and dot $j$ which acquires the Peierl's phase if field $\mathbf{B}$ is applied perpendicular to the closed-loop structure of a triangular TQD. $V_{ij}$ is the long range Coulomb interaction between dot $i$ and $j$, $U_i$ is the on-site Coulomb interaction of dot $i$, $g$ is the g-factor of the host semiconductor, $\mu_B$ is the Bohr magneton, and $\mathbf{S}_\alpha$ is the spin of the $\alpha$-th electron.  
For the present study, we set $t_{ij} = t = -0.05 Ry^*$, $E_i=0$, $U_i=U=2.0 Ry^*$, and $V_{ij}=V=0.5 Ry^*$ as the initial state of the isolated triangular TQD system.  $Ry^* = 5.97$ meV is the effective Rydberg in GaAs.  For linear TQD case, we set $t_{13}=0$ and $V_{13} = V/2$.
In this analysis, we assume $E_i$ and $t_{ij}$ are independently tunable parameters that will be varied to generate Berry's phase. Fig.[\ref{fig:TQD}a] and Fig.[\ref{fig:TQD}b] show the schematic picture of a triangular and a linear TQD, respectively.

As discussed in  Ref.~\onlinecite{korkusinski_gimenez_prb2007}, the two arrangements of a TQD lead to two topologically different Hamiltonians.  The low energy spectrum (4 spin-3/2 and 4 spin-1/2 states) of a half-filled TQD is mapped onto the Heisenberg model \cite{scarola_dassarma_prl2004,scarola_dassarma_pra2005,hsieh_hawrylak_prb2010} with one localized spin in each dot,
\begin{equation}
  \label{eq:heisenbergH}
  H = \sum_{i<j} J_{ij} \mathbf{S}_i \cdot \mathbf{S}_j +\sum_i g\mu_B\mathbf{S}_i \cdot \mathbf{B} + \sum_{i<j<k} \chi_{ijk} \mathbf{S}_i \cdot \left( \mathbf{S}_j \times \mathbf{S}_k \right),
\end{equation}
where the exchange interactions $J_{ij}$'s can be derived from the Hubbard model and expressed by microscopic parameters as,

\begin{equation}
  \label{eq:exchangeJ}
  J_{ij} = 2 |t_{ij}|^2 \left( \frac{1}{U - V + (E_i - E_j)} + \frac{1}{U - V - (E_i - E_j)} \right).
\end{equation}  

The coefficient $\chi_{ijk}$ for the chirality operator in Eq.(\ref{eq:heisenbergH}) is non-zero only for a triangular TQD in the presence of a perpendicular magnetic field and is much smaller than $J_{ij}$ .

Since the Heisenberg Hamiltonian commutes with the total $S_y$ operator of the system, we focus on the $S_y = - 1/2$ subspace.  We further assume that an in-plane magnetic field $B_y$ has been applied to separate the $S_y = -1/2$ and $S_y=1/2$ subspaces by the Zeeman energy.    In the $S_y=-1/2$ subspace, a  resonant triangular TQD with all $J_{ij} = J_0$  has the following three eigenstates\cite{hawrylak_korkusinski_ssc2005,korkusinski_gimenez_prb2007,hsieh_hawrylak_prb2010} ,
\begin{align}
  \label{eq:chirbasis}
  \ket{q_+} &= \frac{1}{\sqrt{3}} \left( \ket{\downarrow\downarrow\uparrow}+e^{i\frac{2\pi}{3}}\ket{\downarrow\uparrow\downarrow}+e^{i\frac{4\pi}{3}}\ket{\uparrow\downarrow\downarrow}\right), \\
   \ket{q_-} &= \frac{1}{\sqrt{3}} \left( \ket{\downarrow\downarrow\uparrow}+e^{-i\frac{2\pi}{3}}\ket{\downarrow\uparrow\downarrow}+e^{-i\frac{4\pi}{3}}\ket{\uparrow\downarrow\downarrow}\right), \\
 \ket{S_{3/2}} &= \frac{1}{\sqrt{3}} \left( \ket{\downarrow\downarrow\uparrow}+\ket{\downarrow\uparrow\downarrow}+\ket{\uparrow\downarrow\downarrow}\right).
   \end{align}
 where spin configurations, like  $\ket{\downarrow\downarrow\uparrow}$, label the electron spins in quantum dots from spin up in dot 1 (rightmost) to spin down in dot 3 (leftmost).
The two chiral states, $\ket{q_+}$ and $\ket{q_-}$, constitute the two levels of a chirality-based coded qubit in a triangular TQD.   The coherent manipulation of a coded qubit is achieved by tuning the exchange interactions $J_{ij}$'s.  In the basis of the $\{ \ket{q_{\pm}} \}$, the Heisenberg Hamiltonian with arbitrary $J_{ij}$'s reads,
\begin{equation}
  \label{eq:3x3matrix}
  H^{qb}_{tri} = \begin{bmatrix}
    0 &   \frac{1}{4}(2J_{23}-J_{13}-J_{12}) - i\frac{\sqrt{3}}{4}(J_{13}-J_{12}) \\
    \frac{1}{4}(2J_{23}-J_{13}-J_{12})+i\frac{\sqrt{3}}{4}(J_{13}-J_{12})  & 0 
   \end{bmatrix}.
\end{equation}
This matrix shows that the chirality-based coded qubit manipulated by $J_{ij}$'s is equivalent to spin-1/2 particle under an effective magnetic field in the $x-y$ plane.

For the case of a linear TQD where $J_{12} = J_{23}=J_0$ and $J_{13}=0$, the three eigenstates of the system are\cite{hawrylak_korkusinski_ssc2005,korkusinski_gimenez_prb2007,hsieh_hawrylak_prb2010}
\begin{align}
  \label{eq:jacobibasis}
  \ket{L_0} &= \frac{1}{\sqrt{2}} \left( \ket{\downarrow\downarrow\uparrow}-\ket{\uparrow\downarrow\downarrow}\right), \\
   \ket{L_1} &= \frac{1}{\sqrt{3}} \left( \ket{\downarrow\downarrow\uparrow}-2\ket{\downarrow\uparrow\downarrow}+\ket{\uparrow\downarrow\downarrow}\right), \\
   \ket{S_{3/2}} &= \frac{1}{\sqrt{3}} \left( \ket{\downarrow\downarrow\uparrow}+\ket{\downarrow\uparrow\downarrow}+\ket{\uparrow\downarrow\downarrow}\right).
 \end{align}
 For the linear TQD, the two Jacobi states, $\ket{L_0}$ and $\ket{L_1}$ constitute the levels of a coded qubit in a linear TQD.   The two Jacobi states used for the coded qubit may be physically distinguished by the joint spin states on dot 1 and dot 3.  For $\ket{L_0}$, the spins in dot 1 and dot 3  form a spin singlet; while the spins in dot 1 and dot 3  form a linear combination of spin triplets with $S_y=-1$ and $S_y=0$ in $\ket{L_1}$.  Similar to the coded qubit in triangular TQD, we present the Heisenberg Hamiltonian of a linear TQD with arbitrary $J_{ij}$'s in the basis of $\{ \ket{L_{0(1)}} \}$,
 \begin{equation}
  \label{eq:3x3matrix2}
  H^{qb}_{lin} = \begin{bmatrix}
    \frac{1}{4} \left( J_{12} + J_{23} \right) & \frac{\sqrt{3}}{4} \left( J_{23} - J_{12} \right) \\
    \frac{\sqrt{3}}{4} \left( J_{23} - J_{12} \right)  & -\frac{1}{4} \left( J_{12} + J_{23} \right)
   \end{bmatrix}.
\end{equation}
 Different from the chirality-based coded qubit, the Jacobi states-defined coded qubit in a linear TQD is equivalent to a  spin-1/2 particle under an effective field in the $x-z$ plane.  Finally, we remark that this Jacobi basis of a linear TQD also diagonalizes  resonant triangular TQD Hamiltonian.  In a triangular TQD, the ground state is doubly degenerate, so  the two Jacobi states, $\ket{L_0}$ and $\ket{L_1}$, are related to the chirality states $\ket{q_\pm}$ via a unitary transformation.

\section{Generation and detection of Berry's phase} \label{sec:generation}

In the previous section, we  presented the Hamiltonian for a coded qubit in both chirality basis (triangular TQD) and Jacobi basis (linear TQD), and compared the coded qubit to a spin-1/2 particle, whose Hamiltonian, $H(\mathbf{R})$, is a function of an effective field, $\mathbf{R} = (X,Y,Z)$, expressed in terms of exchange couplings $J_{ij}$.  
By adiabatically varying the direction of the magnetic field around a closed circuit that encircles the diabolical point
 \cite{berry_wilkinson_rslA1984,berry_frs_rslA1984,herzberg_longuethiggins_df1963}in the parameter space of $\mathbf{R}$, the system undergoes a cyclic evolution and accumulates Berry's phase.  The diabolical point is the point at which the two-level system is degenerate in the parameter space of $\mathbf{R}$.

\begin{figure}
  \includegraphics[width=0.8\textwidth, angle=270]{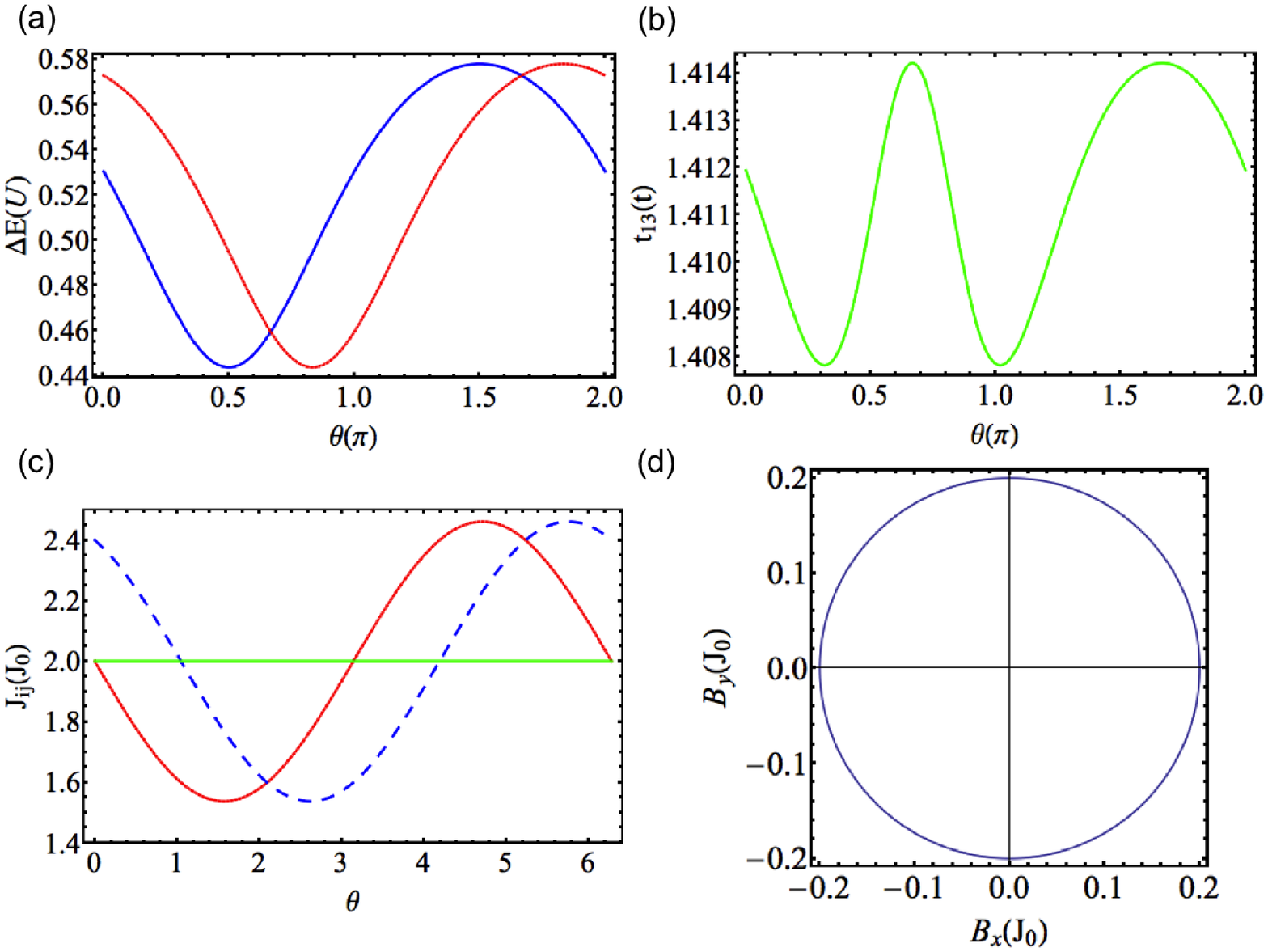}
  \caption{\label{fig:Herzberg1}(Color online) (a) Variations of $\Delta E_{12}$ (blue curve) and $\Delta E_{23}$ (red curve) over one round on the closed circuit.  Note that the variations is comparable to $U$, the largest energy scale of the Hubbard model. (b) Variation of $t_{13}$ over one round on the closed circuit. (c) The values of $J_{ij}$, computed with Eq.(\ref{eq:exchangeJ}), over one round on the closed circuit. $J_{12}$ is red, $J_{23}$ is blue, and $J_{13}$ is green.  This circuit is generated under the constraint that $J_
{13}$ is held constant.  (d) Parametric plot of the closed circuit itself in the parameter space $\mathbf{R}$ for the coded qubit (effective two-level system).  The $x$ and $y$ components of the circle are related to the $J_{ij}$ via the real and imaginary components of the off-diagonal matrix element in Eq.(\ref{eq:3x3matrix}). }
\end{figure}

\begin{figure}
  \includegraphics[width=0.95\textwidth]{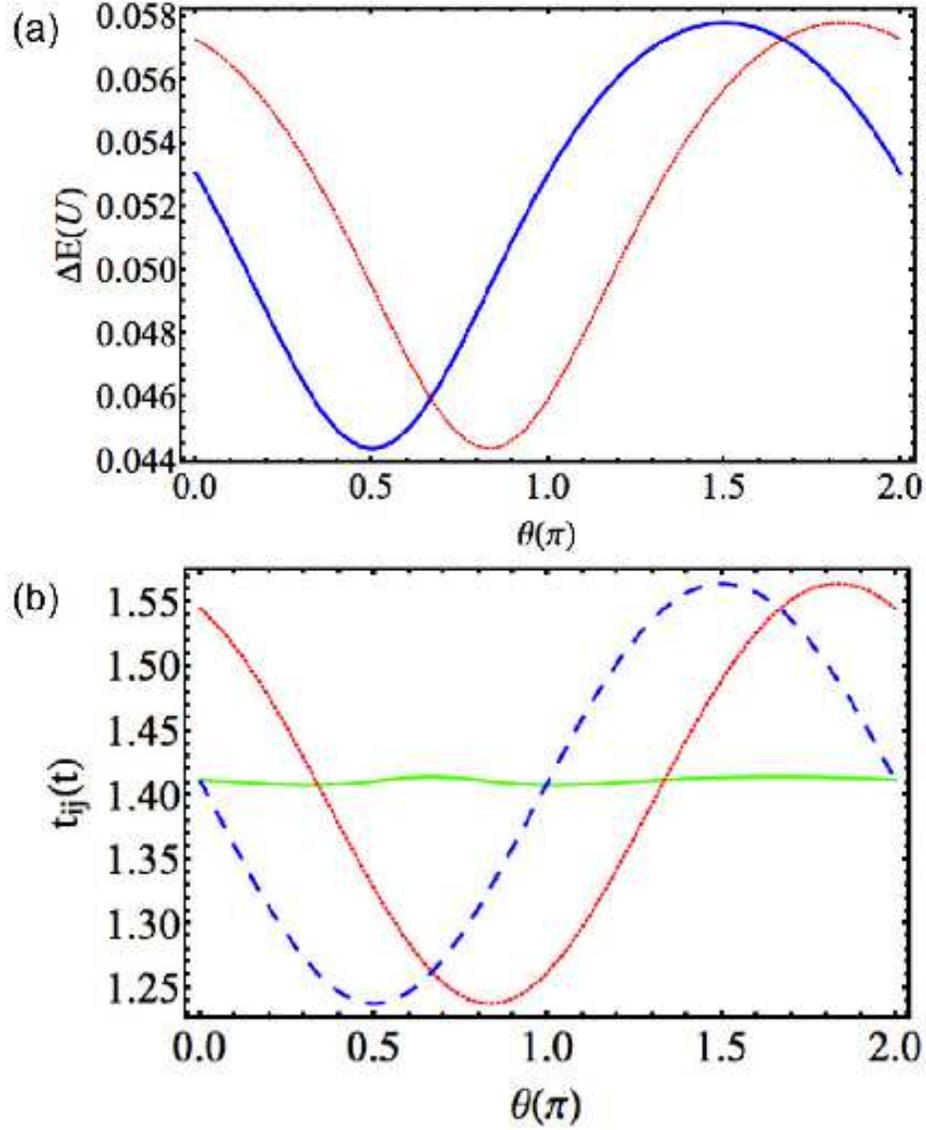}
  \caption{\label{fig:Herzberg2}(Color online) (a) Variations of $\Delta E_{12}$ (blue) and $\Delta E_{23}$ (red) over one round on the closed circuit. The variations of $\Delta E_{ij}$ are significantly suppressed when compared to Fig.[\ref{fig:Herzberg1}a].  This is because we vary all $t_{ij}$.  (b) Variation of $t_{ij}$ over one round on the closed circuit. $t_{12}$ is blue, $t_{23}$ is red, and $t_{13}$ is green.  The values of $t_{ij}$ are chosen specifically to reproduce the same $J_{ij}$ in Fig.[\ref{fig:Herzberg1}c] and  reduce the variations of $\Delta E_i$.}
\end{figure}

\begin{figure}
  \includegraphics[width=0.9\textwidth]{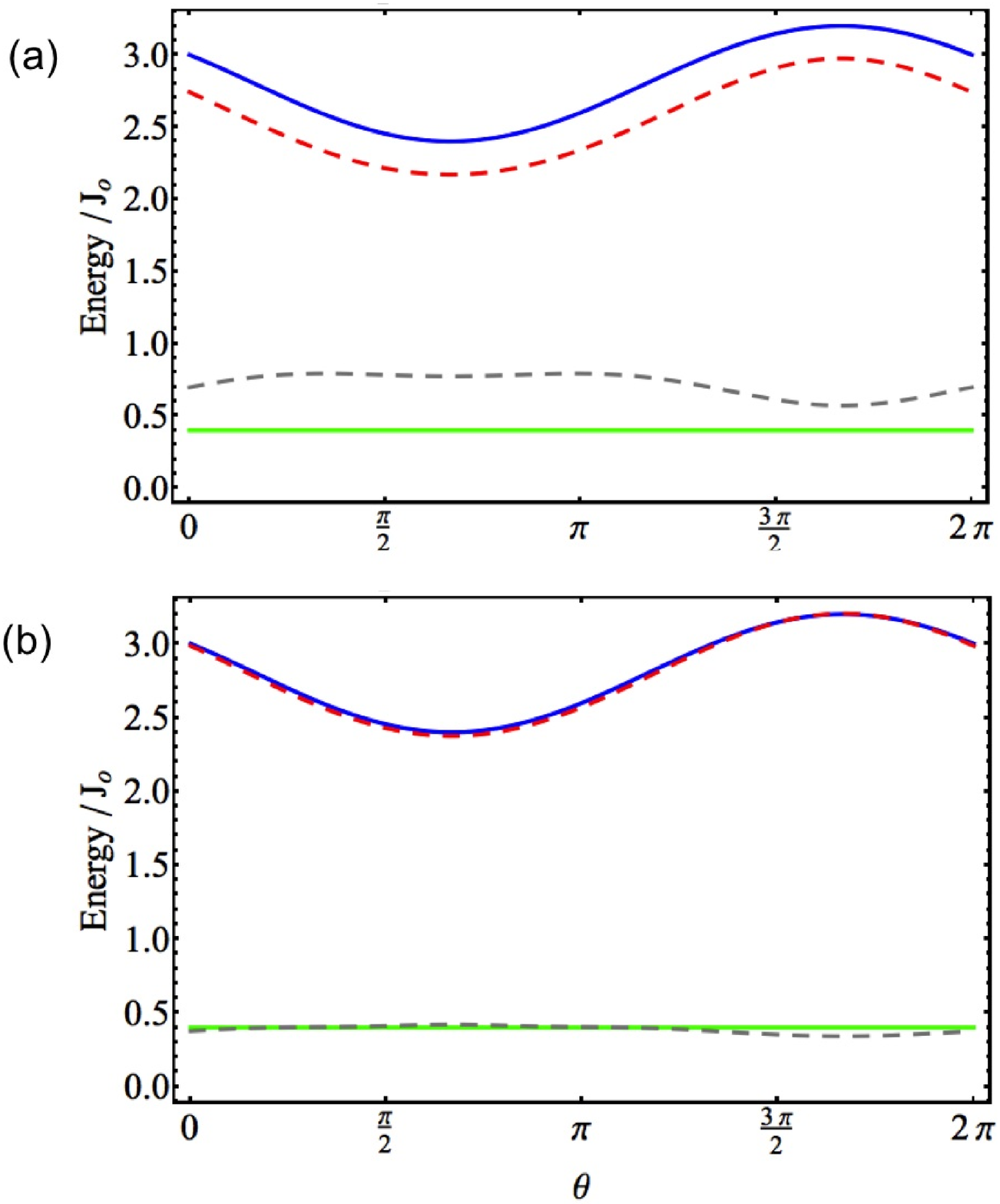}
  \caption{\label{fig:HubbvsHeis}(Color online) (a) The energy gap $\epsilon_{2}-\epsilon_{1}$ (higher ones in the plot)  as well as $\epsilon_{1} - \epsilon_{0}$ (lower ones in the plot) in both Hubbard (dotted curve) and Heisenberg (solid curve) model for the circuit presented in Fig.[\ref{fig:Herzberg1}]. $\epsilon_i$ is the i-th eigenenergy of the system.  (b) The same energy gaps  in both Hubbard (dotted) and Heisenberg (solid) model for the circuit presented in Fig.[\ref{fig:Herzberg2}].  In both (a) and (b), the green, solid curve which represents the energy gap between the two coded qubit level is a constant over the closed circuit.  This is because we transport the coded qubit on a constant energy surface as implied by the parametric plot of $\{B_x , B_y\}$ for the coded qubit (effective two-level system) in Fig.[\ref{fig:Herzberg1}d].  }
\end{figure}

To generate Berry's phase, we need to vary exchange couplings to rotate the effective magnetic field $\mathbf{R}$.
We note that the real and imaginary part of the off-diagonal matrix element in Eq.(\ref{eq:3x3matrix}) can be independently set to either positive or negative value,  if we assume the capability to independently control all three exchange interactions.  Thus, we can engineer the effective field $\mathbf{R}$ for a chirality-based coded qubit to adiabatically traverse a closed circuit that encircles the origin in the parameter space. The next task is then to tune the exchange interactions with experimentally accessible quantities, such as $E_i$ and $t_{ij}$.  Since we need to independently control three exchange interactions, we need to vary at least three variables.  Experimentally, it is usually more desirable to tune the on-site energies than the tunnel couplings.   So the simplest attempt is to vary just the on-site energy of each dot.   However, this simple attempt fails, because each exchange interaction $J_{ij}$ depends on the on-site energies of dot $i$ and dot $j$ through the energy difference $\Delta E_{ij} = E_i - E_j$.   We note a constraint $\Delta E_{12} + \Delta E_{23} + \Delta E_{31} = 0$ exists for  these three energy differences.  Therefore, we have actually two degrees of freedom with the three on-site energies.  We need to  tune at least one tunnel coupling.  For instance, the effective field can be rotated in a perfect circle with a radius $J$ by tuning  $E_1$, $E_3$, and  $t_{13}$, while $E_2 = E$, $t_{23}=t_{12}=t$ are being held constant.  Furthermore, we select values of $E_1$, $E_3$, and $t_{13}$ at each point on the circle such that $J_{13} = \frac{4\gamma}{\sqrt{3}} J$  is obeyed at all times, and $\gamma$ is a constant.  Given these conditions, we need to tune three parameters according to
\begin{align}
\label{eq:circuita}
E_1(\theta) & =  E + \sqrt{(U-V)^2 - \frac{{4(U-V)\vert t \vert ^2}}{\frac{J}{\beta}(\gamma - \sin\theta)}}, \\
\label{eq:circuitb}
E_3(\theta) & =  E + \sqrt{(U-V)^2 - \frac{{4(U-V)\vert t \vert ^2}}{ \frac{J}{2\alpha}\cos\theta+\frac{J}{2\beta}(2\gamma - \sin\theta)}}, \\
\label{eq:circuitc}
t_{13}(\theta) & = \sqrt{\frac{((U-V)^2-(E_1-E_3)^2)\gamma J}{4\beta(U-V)}},
\end{align}
where $\alpha = 1/4$, $\beta = \sqrt{3}/4$, and $\theta$ is the accumulated angle on the closed circuit.  Fig.[\ref{fig:Herzberg1}a] and Fig.[\ref{fig:Herzberg1}b] present the variations of the $\Delta E_{12}$, $\Delta E_{23}$ and $t_{13}$ as functions of $\theta$, and Fig.[\ref{fig:Herzberg1}c] shows the variations of $J_{ij}$ in response to the changes in the Hubbard parameters.  Fig.[\ref{fig:Herzberg1}d] shows the Herzberg circuit with radius $J$ in the parametric space $\mathbf{R}$ for the effective spin-1/2 model.  Although the circuit presented in Fig.[\ref{fig:Herzberg1}] indeed encloses the diabolical point in the parameter space, it is based on the Heisenberg model for an effective two-level system.  The two lowest states in the $S_y=-1/2$ subspace constitute a two-level system modelled by Heisenberg Hamiltonian only when on-site Coulomb repulsion is the dominant energy scale.  If we examine  the scale of variations for $\Delta E_{13}$ and $\Delta E_{23}$ in Fig.[\ref{fig:Herzberg1}], the variations are as high as $0.58$ U at certain points on the circuit.  The validity of the Heisenberg model becomes questionable over the course of transporting the system around the circuit.  Thus, we expect that a more realistic circuit, which minimizes the variations of $\Delta E_{ij}$ in order to ensure the validity of Heisenberg model, would require to tune more than just three variables.  For instance, Fig.[\ref{fig:Herzberg2}a] and Fig.[\ref{fig:Herzberg2}b] present another circuit, which vary again on-site energy $E_1$ and $E_3$ but all $t_{ij}$ in order to produce identical $J_{ij}$ as shown in Fig.[\ref{fig:Herzberg1}c] and also the same circle in Fig.[\ref{fig:Herzberg1}d].  By  moderately varying $t_{12}$ and $t_{23}$, we observe that the variations of $\Delta E_{ij}$ are significantly suppressed as shown in Fig.[\ref{fig:Herzberg2}a].  Fig.[\ref{fig:HubbvsHeis}a] and Fig.[\ref{fig:HubbvsHeis}b] compares the energy gap between the lowest three levels in the Hubbard model and the Heisenberg model in the $S_y=-1/2$ subspace for circuit presented in Fig.[\ref{fig:Herzberg1}] and Fig.[\ref{fig:Herzberg2}], respectively.  As shown,  the second circuit which varies all three tunnel couplings provides a better agreement between the Hubbard and Heisenberg model.  For many tasks, such as manipulating the quantum state of a coded qubit via dynamical phases, a fast tuning of on-site energies is preferred to the tuning of tunnel couplings.  However, for the adiabatic accumulation of geometrical phases  the gating operations can be done at a slower rate, with the tuning of the tunnel couplings  more favourable in this case.

\begin{figure}
  \includegraphics[width=0.75\textwidth, angle=270]{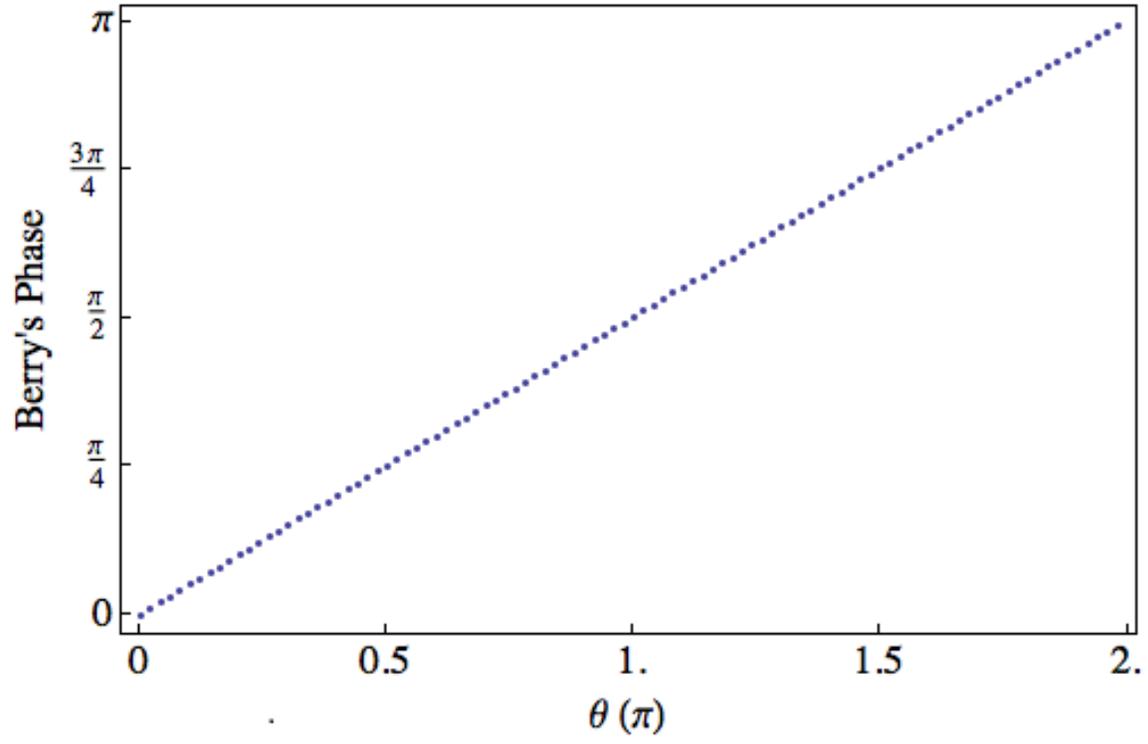}
  \caption{\label{fig:berryphase} Accumulation of Berry's phase by the coded qubit level, $\ket{q_+}$.  At the end of the Herzberg circuit, $\theta = 2 \pi$, the accumulated phase is $\pi$, in agreement with theory. }
\end{figure}

The accumulated Berry's phase, $\phi_{\pm} (C)$, for the coded qubit level $\ket{q_{\pm}}$, after one round along the closed circuit $C$  enclosing the origin is simply $\phi_{\pm} (C) =  \mp i\frac{1}{2}\Omega(C) $, where $\Omega(C)$ is the solid angle subtended by the closed circuit $C$ with respect to the origin of the parameter space.   For the chirality-based coded qubit, the effective field $\mathbf{R}$ is restricted to lie in the $x-y$ plane as implied by Eq.(\ref{eq:3x3matrix}).  The solid angle subtended by any closed circuit in the $x-y$ plane of the parameter space is either $0$ or $\pi$, and this result depends solely on whether the closed circuit encircles the origin (diabolical point) or not.  Fig.[\ref{fig:berryphase}] shows the numerical computation of accumulated geometrical phase for the coded qubit along the Herzberg circuit presented in Fig.[\ref{fig:Herzberg1}].  After one round on the circuit, the state accumulates a phase of $\pi$ in agreement with the theory.  To get Fig.[\ref{fig:berryphase}], we simulate the time evolution, $\psi(\theta)$, of the coded qubit by varying the on-site energies  and tunnel coupling according to Eq.(\ref{eq:circuita}) - Eq.(\ref{eq:circuitc}). We use the angle $\theta$, which denotes the fraction of the Herzberg circuit, as a time variable.  Next, we use the definition of geometrical phase \cite{berry_frs_rslA1984},
\begin{equation}
\label{eq:berrywavefn}
\psi(\theta) = Te^{-\frac{i}{\hbar}\int d\theta' E_{n}(\mathbf{R}(\theta))} e^{i\gamma(\theta)} \ket{n({\mathbf{R}}(\theta))},
\end{equation}
where $ E_{n}(\mathbf{R}(\theta))$ and $\ket{n({\mathbf{R}}(\theta))}$ are the n-th  eigenenergy and eigenfunction for the Hamiltonian with parameter $\mathbf{R}$. We remark that $\ket{n(\mathbf{R}(\theta))}$ should be chosen to be single-valued and the phases of the eigenstates at different $\mathbf{R}$ should be continuously differentiable with respect to $\mathbf{R}$. By Eq.(\ref{eq:berrywavefn}), one may extract the accumulated Berry's phase in a numerical calculation. 

 Let us now turn to the experimental observation of  Berry's phase using quantum interference.
The restriction of having an effective in-plane magnetic field $\mathbf{R}=(X,Y)$ makes the experimental probing of Berry's phase difficult because we have very limited interference between components of the wavefunction.  Therefore, it is  essential to have a closed circuit that  lies above the $x-y$ plane so the solid angle subtended by the circuit is given by $\Omega = 2\pi (1-\cos\Theta)$, where $\cos\Theta$ is the direction cosine of the effective field along the $z$ direction.  To generate an effective field along the $z$ direction, we  need to apply a magnetic field perpendicular to the triangular TQD.  The perpendicular field  turns on the chiral term in Eq.(\ref{eq:heisenbergH}).  Since the chiral states are eigenstates of the chirality operator, the chirality term acts as a $\sigma_z$ operation in the computational space.  The coefficient, $\chi_{ijk}$, attached to the chirality operator is  $ \approx t^3/U^2$.  If we take $t=-0.05$ and $U=2.0$, the values used to generate Fig.[\ref{fig:Herzberg2}], then $\chi_{ijk}$ is about two to three orders of magnitude smaller than  $J_{ij}$'s.  It is important to realize that $\chi_{ijk}$, derived from third order perturbation theory, also depends on the TQD parameters such as $t_{ij}$, and differences of on-site energies through terms like $\frac{1}{(U-V+\Delta E_{ij})(U-V+\Delta E_{kl})}$.  Therefore, as the system is adiabatically transported on a closed circuit by varying the gate voltage, the magnitude of this effective $z$ field will oscillate in its magnitude.  Furthermore, an estimated  change of the effective $z$ field over the course of one complete circuit is  two to three orders of magnitude smaller than due to the $J_{ij}$'s.

With all the necessary ingredients in place, we now describe a possible procedure to experimentally detect Berry's phase in a triangular TQD.  First, we  prepare the coded qubit in a linear superposition of the form $\frac{1}{\sqrt{2}} \ket{q_+} + \frac{1}{\sqrt{2}}  \ket{q_-} $ by having an effective field $\mathbf{R}$ along the $y$ direction.  Next, we  follow Eqs.(\ref{eq:circuita})-(\ref{eq:circuitc}) to tune $E_1$, $E_3$, and $t_{13}$ under the perpendicular magnetic field to accumulate the geometrical phase over one closed circuit lying above the $x-y$ plane.  Over the course of transporting the system around the circuit, the system  acquires both dynamical and geometrical phases.  Therefore, we next have to perform a spin-echo \cite{falci_fazio_nat2000,leek_flink_sci2007,jones_vedral_nat2000} procedure to eliminate the dynamical phases.  For this spin-echo procedure, we perform a NOT operation on the  coded qubit to flip the two chiral states.  Then we transport the system around the same circuit  in the opposite direction.  In this way, the dynamical phases will add destructively while the geometrical phases will add constructively.  At the end of the second round along the circuit, the system now has only Berry's phase left.  The final task is to perform quantum interferometry to extract Berry's phase from the system.   At this stage, it is important to recall the Jacobi states, $\ket{L_0}$ and $\ket{L_1}$, which can be distinguished by the joint spin states on dot 1 and dot 3.  We remark that experiments \cite{petta_johnson_sci2005,koppens_buizert_nat2006} have already demonstrated that the joint spin states between two neighbouring dots can be measured by the spin blockade phenomenon.  If the charge detection shows that we can make a transition from $(1,1,1)$ to $(2,1,0)$, then we have a $\ket{L_0}$ state in the triangular TQD.  Therefore, by simply projecting the chirality-based coded qubit onto the Jacobi states, the probability of detecting $\ket{L_0}$ state is given by
\begin{equation}
\label{eq:measure}
P(L_0) = \frac{3}{2}\cos2\Omega - \frac{\sqrt{3}}{2}\sin2\Omega.
\end{equation}
As mentioned in the previous section, the two bases are related by a unitary transformation.  The measurements in the Jacobi basis thus implicitly provide the needed quantum interference to extract Berry's phase.

We now focus on the case of the coded qubit in a linear TQD.  We note that the diagonal element $H_{11}$ of the Hamiltonian, Eq.(\ref{eq:3x3matrix2}), is always  positive  because all $J_{ij} > 0$, a condition dictated by the microscopic details given in Eq.(\ref{eq:exchangeJ}), in which $U$ is the largest energy scale.  This  implies that when we map the coded qubit in a linear TQD onto an effective spin-1/2, the effective magnetic field component in the $z$ direction is always positive. Hence effective magnetic field in the $z-x$ plane cannot complete a circuit enclosing the orgin of the parameter, i.e., effective magnetic field, space.  Hence we conclude that it is not possible to generate Herzberg circuit and Berry's phase with a coded qubit in a linear TQD.

\section{Conclusion} \label{sec:conclusion}

In summary, we presented a theoretical proposal for the  Herzberg circuit and controlled accumulation of Berry's phase in a  qubit encoded in the two degenerate chirality states of a three spin complex with total spin $S=1/2$ in a triangular triple quantum dot molecule with one electron spin each.   Using a Hubbard and Heisenberg model the Herzberg circuit encircling the degeneracy point is realized by adiabatically tuning the successive on-site energies of quantum dots and tunnel couplings across pairs of neighbouring dots.  It is explicitly shown that encircling the degeneracy point leads to the accumulation of the geometrical Berrys phase. We show that only triangular but not linear quantum dot molecule allows for the generation of Berry's phase and we discuss a protocol to detect this geometrical phase in interference experiment relying on spin blockade spectroscopy. 

\section{Acknowledgment} \label{sec:Ack}
The authors thank  CIFAR , NSERC, QUANTUMWORKS, and  OGS for financial support.


\begin{thebibliography}{24}
\expandafter\ifx\csname natexlab\endcsname\relax\def\natexlab#1{#1}\fi
\expandafter\ifx\csname bibnamefont\endcsname\relax
  \def\bibnamefont#1{#1}\fi
\expandafter\ifx\csname bibfnamefont\endcsname\relax
  \def\bibfnamefont#1{#1}\fi
\expandafter\ifx\csname citenamefont\endcsname\relax
  \def\citenamefont#1{#1}\fi
\expandafter\ifx\csname url\endcsname\relax
  \def\url#1{\texttt{#1}}\fi
\expandafter\ifx\csname urlprefix\endcsname\relax\def\urlprefix{URL }\fi
\providecommand{\bibinfo}[2]{#2}
\providecommand{\eprint}[2][]{\url{#2}}

\bibitem[{\citenamefont{Herzberg and
  Longuet-Higgins}(1963)}]{herzberg_longuethiggins_df1963}
\bibinfo{author}{\bibfnamefont{G.}~\bibnamefont{Herzberg}} \bibnamefont{and}
  \bibinfo{author}{\bibfnamefont{H.}~\bibnamefont{Longuet-Higgins}},
  \bibinfo{journal}{Discuss. Faraday Soc.} \textbf{\bibinfo{volume}{35}},
  \bibinfo{pages}{77} (\bibinfo{year}{1963}).

\bibitem[{\citenamefont{Berry and S}(1984)}]{berry_frs_rslA1984}
\bibinfo{author}{\bibfnamefont{M.}~\bibnamefont{Berry}} \bibnamefont{and}
  \bibinfo{author}{\bibfnamefont{F.~R.} \bibnamefont{S}},
  \bibinfo{journal}{Proc. R. Soc. Lond. A} \textbf{\bibinfo{volume}{392}},
  \bibinfo{pages}{45} (\bibinfo{year}{1984}).

\bibitem[{\citenamefont{Wilczek and Shapere}(1988)}]{wilczek_shapere_book}
\bibinfo{author}{\bibfnamefont{F.}~\bibnamefont{Wilczek}} \bibnamefont{and}
  \bibinfo{author}{\bibfnamefont{A.}~\bibnamefont{Shapere}},
  \emph{\bibinfo{title}{Geometric Phases in Physics}}
  (\bibinfo{publisher}{Wolrd Scientific}, \bibinfo{year}{1988}).

\bibitem[{\citenamefont{Carollo and Vedral}(2008)}]{carollo_vlatko_book2008}
\bibinfo{author}{\bibfnamefont{A.~C.} \bibnamefont{Carollo}} \bibnamefont{and}
  \bibinfo{author}{\bibfnamefont{V.}~\bibnamefont{Vedral}},
  \emph{\bibinfo{title}{Holonomic Quantum Computation}}
  (\bibinfo{publisher}{Wiley-VCH Verlag GmbH}, \bibinfo{year}{2008}), pp.
  \bibinfo{pages}{381--387}.

\bibitem[{\citenamefont{Jiannis et~al.}(1999)\citenamefont{Jiannis, Paolo, and
  Mario}}]{pachos_zanardi_pra2000}
\bibinfo{author}{\bibfnamefont{P.}~\bibnamefont{Jiannis}},
  \bibinfo{author}{\bibfnamefont{Z.}~\bibnamefont{Paolo}}, \bibnamefont{and}
  \bibinfo{author}{\bibfnamefont{R.}~\bibnamefont{Mario}},
  \bibinfo{journal}{Phys. Rev. A} \textbf{\bibinfo{volume}{61}},
  \bibinfo{pages}{010305} (\bibinfo{year}{1999}).

\bibitem[{\citenamefont{Bitter and Dubbers}(1987)}]{bitter_dubbers_prl1987}
\bibinfo{author}{\bibfnamefont{T.}~\bibnamefont{Bitter}} \bibnamefont{and}
  \bibinfo{author}{\bibfnamefont{D.}~\bibnamefont{Dubbers}},
  \bibinfo{journal}{Phys. Rev. Lett.} \textbf{\bibinfo{volume}{59}},
  \bibinfo{pages}{251} (\bibinfo{year}{1987}).

\bibitem[{\citenamefont{Jones et~al.}(2000)\citenamefont{Jones, Vedral, Ekert,
  and Castagnoli}}]{jones_vedral_nat2000}
\bibinfo{author}{\bibfnamefont{J.~A.} \bibnamefont{Jones}},
  \bibinfo{author}{\bibfnamefont{V.}~\bibnamefont{Vedral}},
  \bibinfo{author}{\bibfnamefont{A.}~\bibnamefont{Ekert}}, \bibnamefont{and}
  \bibinfo{author}{\bibfnamefont{G.}~\bibnamefont{Castagnoli}},
  \bibinfo{journal}{Nature} \textbf{\bibinfo{volume}{403}}, \bibinfo{pages}{1}
  (\bibinfo{year}{2000}).

\bibitem[{\citenamefont{Leek et~al.}(2007)\citenamefont{Leek, Fink, Blais,
  Bianchetti, G{\"o}ppl, Gambetta, Schuster, Frunzio, Schoelkopf, and
  Wallraff}}]{leek_flink_sci2007}
\bibinfo{author}{\bibfnamefont{P.~J.} \bibnamefont{Leek}},
  \bibinfo{author}{\bibfnamefont{J.~M.} \bibnamefont{Fink}},
  \bibinfo{author}{\bibfnamefont{A.}~\bibnamefont{Blais}},
  \bibinfo{author}{\bibfnamefont{R.}~\bibnamefont{Bianchetti}},
  \bibinfo{author}{\bibfnamefont{M.}~\bibnamefont{G{\"o}ppl}},
  \bibinfo{author}{\bibfnamefont{J.~M.} \bibnamefont{Gambetta}},
  \bibinfo{author}{\bibfnamefont{D.~I.} \bibnamefont{Schuster}},
  \bibinfo{author}{\bibfnamefont{L.}~\bibnamefont{Frunzio}},
  \bibinfo{author}{\bibfnamefont{R.~J.} \bibnamefont{Schoelkopf}},
  \bibnamefont{and} \bibinfo{author}{\bibfnamefont{A.}~\bibnamefont{Wallraff}},
  \bibinfo{journal}{Science} \textbf{\bibinfo{volume}{318}},
  \bibinfo{pages}{1889} (\bibinfo{year}{2007}).

\bibitem[{\citenamefont{M{\"o}tt{\"o}nen
  et~al.}(2008)\citenamefont{M{\"o}tt{\"o}nen, Vartiainen, and
  Pekola}}]{mottonen_vartiainen_prl2008}
\bibinfo{author}{\bibfnamefont{M.}~\bibnamefont{M{\"o}tt{\"o}nen}},
  \bibinfo{author}{\bibfnamefont{J.~J.} \bibnamefont{Vartiainen}},
  \bibnamefont{and} \bibinfo{author}{\bibfnamefont{J.~P.}
  \bibnamefont{Pekola}}, \bibinfo{journal}{Phys. Rev. Lett.}
  \textbf{\bibinfo{volume}{100}}, \bibinfo{pages}{177201}
  (\bibinfo{year}{2008}).

\bibitem[{\citenamefont{Aunola and Toppari}(2003)}]{aunola_toppari_prb2003}
\bibinfo{author}{\bibfnamefont{M.}~\bibnamefont{Aunola}} \bibnamefont{and}
  \bibinfo{author}{\bibfnamefont{J.~J.} \bibnamefont{Toppari}},
  \bibinfo{journal}{Phys. Rev. B} \textbf{\bibinfo{volume}{68}},
  \bibinfo{pages}{020502(R)} (\bibinfo{year}{2003}).

\bibitem[{\citenamefont{Pekola et~al.}(1999)\citenamefont{Pekola, Toppari,
  Aunola, Savolainen, and Averin}}]{pekola_toppari_prb1999}
\bibinfo{author}{\bibfnamefont{J.~P.} \bibnamefont{Pekola}},
  \bibinfo{author}{\bibfnamefont{J.~J.} \bibnamefont{Toppari}},
  \bibinfo{author}{\bibfnamefont{M.}~\bibnamefont{Aunola}},
  \bibinfo{author}{\bibfnamefont{M.~T.} \bibnamefont{Savolainen}},
  \bibnamefont{and} \bibinfo{author}{\bibfnamefont{D.~V.}
  \bibnamefont{Averin}}, \bibinfo{journal}{Phys. Rev. B}
  \textbf{\bibinfo{volume}{60}}, \bibinfo{pages}{R9931} (\bibinfo{year}{1999}).

\bibitem[{\citenamefont{Falci et~al.}(2000)\citenamefont{Falci, Fazio, Palma,
  Siewert, and Vedral}}]{falci_fazio_nat2000}
\bibinfo{author}{\bibfnamefont{G.}~\bibnamefont{Falci}},
  \bibinfo{author}{\bibfnamefont{R.}~\bibnamefont{Fazio}},
  \bibinfo{author}{\bibfnamefont{G.~M.} \bibnamefont{Palma}},
  \bibinfo{author}{\bibfnamefont{J.}~\bibnamefont{Siewert}}, \bibnamefont{and}
  \bibinfo{author}{\bibfnamefont{V.}~\bibnamefont{Vedral}},
  \bibinfo{journal}{Nature} \textbf{\bibinfo{volume}{407}},
  \bibinfo{pages}{355} (\bibinfo{year}{2000}).

\bibitem[{\citenamefont{Hawrylak and
  Korkusinski}(2005)}]{hawrylak_korkusinski_ssc2005}
\bibinfo{author}{\bibfnamefont{P.}~\bibnamefont{Hawrylak}} \bibnamefont{and}
  \bibinfo{author}{\bibfnamefont{M.}~\bibnamefont{Korkusinski}},
  \bibinfo{journal}{Solid State Communications} \textbf{\bibinfo{volume}{136}},
  \bibinfo{pages}{508} (\bibinfo{year}{2005}).

\bibitem[{\citenamefont{Hsieh and Hawrylak}(2010)}]{hsieh_hawrylak_prb2010}
\bibinfo{author}{\bibfnamefont{C.-Y.} \bibnamefont{Hsieh}} \bibnamefont{and}
  \bibinfo{author}{\bibfnamefont{P.}~\bibnamefont{Hawrylak}},
  \bibinfo{journal}{Phys. Rev. B} \textbf{\bibinfo{volume}{82}},
  \bibinfo{pages}{205311} (\bibinfo{year}{2010}).

\bibitem[{\citenamefont{DiVincenzo et~al.}(2000)\citenamefont{DiVincenzo,
  Bacon, Kempe, and Burkard}}]{divincenzo_bacon_nat2000}
\bibinfo{author}{\bibfnamefont{D.}~\bibnamefont{DiVincenzo}},
  \bibinfo{author}{\bibfnamefont{D.}~\bibnamefont{Bacon}},
  \bibinfo{author}{\bibfnamefont{J.}~\bibnamefont{Kempe}}, \bibnamefont{and}
  \bibinfo{author}{\bibfnamefont{G.}~\bibnamefont{Burkard}},
  \bibinfo{journal}{Nature} \textbf{\bibinfo{volume}{408}},
  \bibinfo{pages}{339} (\bibinfo{year}{2000}).

\bibitem[{\citenamefont{Gaudreau et~al.}(2009)\citenamefont{Gaudreau, Kam,
  Granger, Studenikin, Zawadzki, and Sachrajda}}]{gaudreau_kam_apl2009}
\bibinfo{author}{\bibfnamefont{L.}~\bibnamefont{Gaudreau}},
  \bibinfo{author}{\bibfnamefont{A.}~\bibnamefont{Kam}},
  \bibinfo{author}{\bibfnamefont{G.}~\bibnamefont{Granger}},
  \bibinfo{author}{\bibfnamefont{S.~A.} \bibnamefont{Studenikin}},
  \bibinfo{author}{\bibfnamefont{P.}~\bibnamefont{Zawadzki}}, \bibnamefont{and}
  \bibinfo{author}{\bibfnamefont{A.~S.} \bibnamefont{Sachrajda}},
  \bibinfo{journal}{Appl. Phys. Lett.} \textbf{\bibinfo{volume}{95}},
  \bibinfo{pages}{193101} (\bibinfo{year}{2009}).

\bibitem[{\citenamefont{Granger et~al.}(2010)\citenamefont{Granger, Gaudreau,
  Kam, and Pioro-Ladri{\`e}re}}]{granger_gaudreau_prb2010}
\bibinfo{author}{\bibfnamefont{G.}~\bibnamefont{Granger}},
  \bibinfo{author}{\bibfnamefont{L.}~\bibnamefont{Gaudreau}},
  \bibinfo{author}{\bibfnamefont{A.}~\bibnamefont{Kam}}, \bibnamefont{and}
  \bibinfo{author}{\bibfnamefont{M.}~\bibnamefont{Pioro-Ladri{\`e}re}},
  \bibinfo{journal}{Phys. Rev. B} \textbf{\bibinfo{volume}{82}},
  \bibinfo{pages}{075304} (\bibinfo{year}{2010}).

\bibitem[{\citenamefont{Laird et~al.}(2010)\citenamefont{Laird, Taylor,
  DiVincenzo, Marcus, Hanson, and Gossard}}]{laird_taylor_prb2010}
\bibinfo{author}{\bibfnamefont{E.~A.} \bibnamefont{Laird}},
  \bibinfo{author}{\bibfnamefont{J.~M.} \bibnamefont{Taylor}},
  \bibinfo{author}{\bibfnamefont{D.~P.} \bibnamefont{DiVincenzo}},
  \bibinfo{author}{\bibfnamefont{C.~M.} \bibnamefont{Marcus}},
  \bibinfo{author}{\bibfnamefont{M.~P.} \bibnamefont{Hanson}},
  \bibnamefont{and} \bibinfo{author}{\bibfnamefont{A.~C.}
  \bibnamefont{Gossard}}, \bibinfo{journal}{Phys. Rev. B}
  \textbf{\bibinfo{volume}{82}}, \bibinfo{pages}{075403}
  (\bibinfo{year}{2010}).

\bibitem[{\citenamefont{Korkusinski et~al.}(2007)\citenamefont{Korkusinski,
  Gimenez, Hawrylak, Gaudreau, Studenikin, and
  Sachrajda}}]{korkusinski_gimenez_prb2007}
\bibinfo{author}{\bibfnamefont{M.}~\bibnamefont{Korkusinski}},
  \bibinfo{author}{\bibfnamefont{I.~P.} \bibnamefont{Gimenez}},
  \bibinfo{author}{\bibfnamefont{P.}~\bibnamefont{Hawrylak}},
  \bibinfo{author}{\bibfnamefont{L.}~\bibnamefont{Gaudreau}},
  \bibinfo{author}{\bibfnamefont{S.}~\bibnamefont{Studenikin}},
  \bibnamefont{and}
  \bibinfo{author}{\bibfnamefont{A.}~\bibnamefont{Sachrajda}},
  \bibinfo{journal}{Phys. Rev. B} \textbf{\bibinfo{volume}{75}},
  \bibinfo{pages}{115301} (\bibinfo{year}{2007}).

\bibitem[{\citenamefont{Scarola et~al.}(2004)\citenamefont{Scarola, Park, and
  Das~Sarma}}]{scarola_dassarma_prl2004}
\bibinfo{author}{\bibfnamefont{V.~W.} \bibnamefont{Scarola}},
  \bibinfo{author}{\bibfnamefont{K.}~\bibnamefont{Park}}, \bibnamefont{and}
  \bibinfo{author}{\bibfnamefont{S.}~\bibnamefont{Das~Sarma}},
  \bibinfo{journal}{Phys. Rev. Lett.} \textbf{\bibinfo{volume}{93}},
  \bibinfo{pages}{120503} (\bibinfo{year}{2004}).

\bibitem[{\citenamefont{Scarola and
  Das~Sarma}(2005)}]{scarola_dassarma_pra2005}
\bibinfo{author}{\bibfnamefont{V.~W.} \bibnamefont{Scarola}} \bibnamefont{and}
  \bibinfo{author}{\bibfnamefont{S.}~\bibnamefont{Das~Sarma}},
  \bibinfo{journal}{Phys. Rev. A} \textbf{\bibinfo{volume}{71}},
  \bibinfo{pages}{032340} (\bibinfo{year}{2005}).

\bibitem[{\citenamefont{Berry and Wilkinson}(1984)}]{berry_wilkinson_rslA1984}
\bibinfo{author}{\bibfnamefont{M.}~\bibnamefont{Berry}} \bibnamefont{and}
  \bibinfo{author}{\bibfnamefont{M.}~\bibnamefont{Wilkinson}},
  \bibinfo{journal}{Proc. R. Soc. Lond. A} \textbf{\bibinfo{volume}{392}},
  \bibinfo{pages}{16} (\bibinfo{year}{1984}).

\bibitem[{\citenamefont{Petta et~al.}(2005)\citenamefont{Petta, Johnson,
  Taylor, Laird, Yacoby, Lukin, Marcus, and Hanson}}]{petta_johnson_sci2005}
\bibinfo{author}{\bibfnamefont{J.}~\bibnamefont{Petta}},
  \bibinfo{author}{\bibfnamefont{A.}~\bibnamefont{Johnson}},
  \bibinfo{author}{\bibfnamefont{J.}~\bibnamefont{Taylor}},
  \bibinfo{author}{\bibfnamefont{E.}~\bibnamefont{Laird}},
  \bibinfo{author}{\bibfnamefont{A.}~\bibnamefont{Yacoby}},
  \bibinfo{author}{\bibfnamefont{M.~D.} \bibnamefont{Lukin}},
  \bibinfo{author}{\bibfnamefont{C.~M.} \bibnamefont{Marcus}},
  \bibnamefont{and} \bibinfo{author}{\bibfnamefont{M.~P.}
  \bibnamefont{Hanson}}, \bibinfo{journal}{Science}
  \textbf{\bibinfo{volume}{309}}, \bibinfo{pages}{2180} (\bibinfo{year}{2005}).

\bibitem[{\citenamefont{Koppens et~al.}(2006)\citenamefont{Koppens, Buizert,
  Tielrooij, and Vink}}]{koppens_buizert_nat2006}
\bibinfo{author}{\bibfnamefont{F.}~\bibnamefont{Koppens}},
  \bibinfo{author}{\bibfnamefont{C.}~\bibnamefont{Buizert}},
  \bibinfo{author}{\bibfnamefont{K.}~\bibnamefont{Tielrooij}},
  \bibnamefont{and} \bibinfo{author}{\bibfnamefont{I.}~\bibnamefont{Vink}},
  \bibinfo{journal}{Nature} \textbf{\bibinfo{volume}{442}},
  \bibinfo{pages}{766} (\bibinfo{year}{2006}).

\end{thebibliography}


\newpage

\end{document}